\documentclass[12pt,aps]{revtex4}

\usepackage{graphicx}
\usepackage{latexsym,amssymb,amsmath}

\newcommand{\C}{\mathcal{C}}

\begin{document}

\title{Einstein Relation for Nonequilibrium Steady States}

\author{T. Hanney and M. R. Evans}
%\email{}
\affiliation{School of Physics and Astronomy, University of
Edinburgh, Mayfield Road, Edinburgh, EH9 3JZ, UK}

\date{December 17, 2002}

%\pacs{Pacs codes}

\begin{abstract}
The Einstein relation, relating the steady state fluctuation
properties to the linear response to a perturbation, 
is considered for steady states
of stochastic models with a finite state space .  We
show how an Einstein relation always holds if the steady state
satisfies detailed balance.  More generally, we consider
nonequilibrium steady states where detailed balance does not hold and show how a generalisation of the
Einstein relation may be derived in certain cases.  In particular, for
the asymmetric simple exclusion process 
and a driven diffusive dimer model, the external perturbation
creates and annihilates particles thus breaking the particle
conservation of the unperturbed model.
\end{abstract}

\maketitle


%%%%%%%%%%%%%%%%%%%%%%%%%%%%%%%%%%%%%%%%%%%%%%%%%%%%%%%%%%%%%%%%%%%%%%%%%%%%%  
%%%% ****************************************************************** %%%%%
%%%%%%%%%%%%%%%%%%%%%%%%%%%%%%%%%%%%%%%%%%%%%%%%%%%%%%%%%%%%%%%%%%%%%%%%%%%%%

\section{Introduction}

A guiding principle of equilibrium statistical mechanics is provided
by the fluctuation-dissipation theorem, which relates the 
fluctuation properties of a system in equilibrium to the response
of the system to an external perturbation. Its simplest form -- the
zero-frequency limit -- is the Einstein relation which
relates linear transport coefficients to spontaneous equilibrium fluctuations.
For example the mobility $\mu$ and diffusion constant $\Delta$ of a
Brownian particle interacting with its environment are related by
\begin{gather}
\mu = \beta \Delta \;,
\end{gather}
where $\beta$ is the inverse temperature. This is a deep result with
many implications.  It enables one to reduce the calculation of
response and transport coefficients (mobility, susceptibility, etc.)
to a zero-field calculation. Further, one can formally consider that
spontaneous equilibrium fluctuations are caused by fictitious random
forces. In particular the Einstein relation dictates the forms of
dissipative terms and noise in a phenomenological description of, for
example, dynamic critical behaviour.

A standard derivation of the Einstein relation (in the above form)
comes from the Langevin equation describing Brownian motion. Here, the
relation implies that diffusion is a
direct consequence of the fluctuations of a Brownian particle, and,
conversely, the existence of a frictional drag on the Brownian
particle implies spontaneous equilibrium fluctuations of a random
force. Because of the general applicability of the Einstein relation, one
can make such statements for any transport problem \cite{callen51}, for
example, the existence of an electrical impedance implies spontaneous
fluctuations of the voltage difference across its terminals \cite{nyquist28}.

The most general formulations, applicable to quantum and classical
models of equilibrium statistical mechanics, are due to Kubo
\cite{kubo57,kubo85,forster75},
and Kadanoff and Martin \cite{forster75,kadanoff63}.
The former relies on formal  
perturbation arguments for the linear response of a system near equilibrium;
the latter approach is based
on correlation functions.
These approaches have been
employed to derive the linearised hydrodynamics for the KLS model -- a
driven diffusive system and nonequilibrium generalisation of the Ising
model -- where the steady state satisfies detailed balance and the
applied perturbation is the drive \cite{katz84}.

Some attempts have been made to generalise the fluctuation-dissipation
theorem to models without detailed balance
\cite{eyink96,bertini01, bertini02}. These consider the relationship between
dissipation and spontaneous steady state fluctuations and have found
most success for steady states where the system can be divided into
cells small on the macroscale, in each of which the dynamics satisfy
detailed balance with respect to a local fixed density product
measure. The global steady state can then be written in terms of these
local product measure states. Such steady states are relevant when
boundary driving breaks the global detailed balance of an equilibrium
model. Also, a fluctuation theorem for nonequilibrium steady states
has been demonstrated by Gallavotti and Cohen
\cite{gallavotti95a,lebowitz99}. 
    
However, it is not known in general how to derive an Einstein relation
for steady states of driven nonequilibrium systems where detailed
balance is not satisfied. Since no such guiding principle exists in
nonequilibrium statistical mechanics, we seek here to generalise the
Einstein relation to nonequilibrium steady states of lattice-based
stochastic models governed by a Master equation. 

In what follows, we review in section~\ref{sect:DA} the approach
introduced by Derrida \cite{derrida83} that yields general expressions
for a current and diffusion constant for systems described by a
Master equation.   We consider stochastic models with a finite state
space and use this approach to investigate the existence of Einstein
relations. In section~\ref{sect:DB} we consider steady states where
detailed balance is satisfied and show generally that an Einstein
relation holds. In  section~\ref{sect:NSS}
we consider nonequilibrium steady states where detailed balance
is not obeyed and discuss the conditions under which an Einstein
relation holds. 
We show that the usual Einstein relation holds for
a simple class of models that includes  boundary-driven
symmetric exclusion (diffusing particles with hard-core interaction).
Beyond this class, we consider  bulk-driven
models such as the asymmetric simple exclusion process (ASEP)
and driven dimer diffusion.
In particular, we obtain an Einstein relation for the ASEP
that relates the diffusion constant to the response of
the current to an applied field that creates and annihilates particles. 
We conclude in section~\ref{sect:conc}.

%%%%%%%%%%%%%%%%%%%%%%%%%%%%%%%%%%%%%%%%%%%%%%%%%%%%%%%%%%%%%%%%%%%%%%%%%%%%%  
%%%% ******************************************************************* %%%%
%%%%%%%%%%%%%%%%%%%%%%%%%%%%%%%%%%%%%%%%%%%%%%%%%%%%%%%%%%%%%%%%%%%%%%%%%%%%%

\section{Calculation of the current and diffusion constant}
\label{sect:DA}
In this section, we outline how to obtain general expressions for a
current $J$ and diffusion constant $\Delta$, for stochastic models
defined on a finite state space. The approach
is given in detail by Derrida \textit{et. al.}
\cite{derrida83,derrida93a,derrida97a}; here we just present the key
equations.

We consider a random variable $Y_t$ which increases and
decreases according to specified dynamical events. For example, for a
system of random walkers, if we take $Y_t$ as the total displacement,
then $Y_t$ increases when a walker steps to the
right and decreases when one steps to the left.
In the long-time limit, the mean and variance
of $Y_t$ increase linearly in time, and the
constants of proportionality are interpreted respectively as a current
(sometimes referred to as a velocity)
and a diffusion constant. This is always true for finite systems,
regardless of whether or not the infinite system exhibits anomalous
diffusion i.e. when the mean or variance of $Y_t$ increases
nonlinearly in time. 
 
The moments of $Y_t$ are obtained from $P_t(\C,Y)$,
which is the probability that at time $t$ the system is in a
configuration $\C$ and the random variable takes the value $Y_t = Y$. The time
evolution of $P_t(\C,Y)$ is governed by a Master equation
\begin{align} \label{ME}
\frac{d}{dt} P_t(\C,Y) =  \sum_{\C^\prime} & \left[ \left\{ M_0(\C,\C^\prime)
P_t(\C^\prime,Y) + M_1(\C,\C^\prime) P_t(\C^\prime,Y{-}1) +
M_{-1}(\C,\C^\prime) P_t(\C^\prime,Y{+}1) \right\} \right. \nonumber \\ & -
 \left. \left\{ M_0(\C^\prime,\C) P_t(\C,Y) + M_1(\C^\prime,\C)
P_t(\C,Y) + M_{-1}(\C^\prime,\C) P_t(\C,Y) \right\} \right] \;,
\end{align}
where $M_0(\C,\C^\prime)$ is a transition rate from the configuration
$\C^\prime$ to the configuration $\C$ without contributing to $Y_t$,
$M_1(\C,\C^\prime)$ contains transitions from $\C^\prime$ to $\C$
which increase $Y_t$ by one and $M_{-1}(\C,\C^\prime)$ contains
transitions from $\C^\prime$ to $\C$ which decrease $Y_t$ by
one.  The diagonal elements $M_0(\C,\C)$, $M_1(\C,\C)$ and
$M_{-1}(\C,\C)$ are defined to be zero.
It is straightforward to obtain evolution equations for the first two
moments of $Y_t$, $\langle Y_t \rangle$ and $\langle Y_t^2 \rangle$,
by multiplying both sides of (\ref{ME}) by $Y$ and $Y^2$. The angled
brackets here denote an average over steady state
initial conditions and all histories of the dynamics. 
(The formalism can be extended to include other initial conditions
\cite{derrida95}.)
These evolution equations are expressed in a compact fashion in terms
of $p_t(\C) = \sum_Y P_t(\C,Y)$ and $q_t(\C) =
\sum_Y Y P_t(\C,Y)$: $p_t(\C)$ is the probability of being in
configuration $\C$ at time $t$; $q_t(\C)/p_t(\C)$ is the average value of
$Y_t$ given the system is in configuration $\C$ at time $t$.
These quantities tend toward the asymptotic steady-state forms \cite{derrida83}
\begin{gather} \label{asymptotics}
p_t(\C) \rightarrow p(\C) \qquad \text{and} \qquad q_t(\C) \rightarrow
Jtp(\C) + r(\C)\;,
\end{gather}
in the long-time limit.
% This form for $p_t(\C)$ reflects the fact the
%the system forgets its initial condition and the form for $q_t(\C)$
%implies that $r(\C)/p(\C)$ is the correction to $\langle Y_t \rangle = Jt$
%due to being in configuration $C$.

Here $J$ is a current defined by $J = \partial_t \langle Y_t
\rangle$ and is found, by substituting the asymptotics into the
evolution equation for $\langle Y_t \rangle$, to be  
\begin{gather} \label{J}
J = \sum_{\C,\C^{\prime}} \left[ M_1 (\C,\C^{\prime}) - M_{-1}
(\C,\C^\prime) \right] p(\C^\prime) \;.
\end{gather}

The diffusion constant $\Delta$ associated with the current $J$ is
defined as $\Delta = \partial_t (\langle
Y_t^2 \rangle - \langle Y_t \rangle^2)$. Then substituting the
asymptotics (\ref{asymptotics}) into the evolution equations for
$\langle Y_t \rangle$ and $\langle Y_t^2 \rangle$ yields
\begin{gather} \label{D}
\Delta = \sum_{\C,\C^\prime} \left[ M_1 (\C,\C^{\prime}) + M_{-1}
(\C,\C^\prime) \right] p(\C^\prime) + 2 \sum_{\C,\C^\prime} \left[ M_1
(\C,\C^{\prime}) - M_{-1} (\C,\C^\prime) \right] r(\C^\prime) - 2J
\sum_{\C} r(\C) \;.
\end{gather}

The quantities $p(\C)$ and $r(\C)$ are determined by
substituting the asymptotic forms (\ref{asymptotics}) into the
evolution equations for $p_t(\C)$ and $q_t(\C)$, which are obtained from
(\ref{ME}): $p(\C)$ satisfies
\begin{gather} 
\label{p(C)}
\sum_{\C^\prime} \left[ M(\C,\C^\prime) p(\C^\prime) - 
M(\C^\prime,\C) p(\C) \right] = 0 \;,  
\end{gather}
where the total transition matrix $M=M_0+M_1+M_{{-}1}$, and $r(\C)$
satisfies
\begin{gather}
\label{r(C)}
\sum_{\C^\prime} \left[ M(\C,\C^\prime) r(\C^\prime) - 
M(\C^\prime,\C) r(\C) \right] = J p(\C) - \sum_{\C^\prime} \left[ M_1
(\C,\C^{\prime}) - M_{-1} (\C,\C^\prime) \right] p(\C^\prime)\;.
\end{gather}
Although (\ref{r(C)}) only fixes $r(\C)$ up to a term proportional to
$p(\C)$, this term will cancel from (\ref{D}), so we take without loss
of generality 
\begin{gather} \label{sumr(C)}
\sum_{\C} r(\C) = 0 \;.
\end{gather}

%%%%%%%%%%%%%%%%%%%%%%%%%%%%%%%%%%%%%%%%%%%%%%%%%%%%%%%%%%%%%%%%%%%%%%%%%%%%%  
%%%% ******************************************************************* %%%%
%%%%%%%%%%%%%%%%%%%%%%%%%%%%%%%%%%%%%%%%%%%%%%%%%%%%%%%%%%%%%%%%%%%%%%%%%%%%%

\section{Einstein relation for systems satisfying detailed balance}
\label{sect:DB} 
We now illustrate how one may exploit the approach outlined in Section
\ref{sect:DA} in order to obtain an Einstein relation in the case
where the dynamics obey detailed balance.  

Firstly, the condition under which detailed balance holds is
\begin{gather} \label{DB}
M(\C,\C^\prime) p(\C^\prime) = M(\C^\prime,\C) p(\C)\;.
\end{gather}
Under the assumption that if $Y_t$ increases by one on going from $\C$
to $\C'$, then $Y_t$ decreases by one on going from $\C'$ to $\C$, 
(\ref{DB}) implies 
\begin{align} \label{DB2}
M_1(\C,\C^\prime) p(\C^\prime) &= M_{-1}(\C^\prime,\C) p(\C) \;,\\
\label{DB3}
M_{-1}(\C,\C^\prime) p(\C^\prime) &= M_1(\C^\prime,\C) p(\C) \;.
\end{align}
Using these in equation (\ref{J}) implies that the current
$J=0$. Therefore, from (\ref{r(C)}), we find that $r(\C)$ satisfies
\begin{gather} \label{unpertr}
\sum_{\C^\prime} \left[ M(\C,\C^\prime) r(\C^\prime) -
M(\C^\prime,\C) r(\C) \right] = - \sum_{\C^\prime} \left[ M_1
(\C,\C^{\prime}) - M_{-1} (\C,\C^\prime) \right] p(\C^\prime)\;.
\end{gather}  

Next, we consider a perturbation which couples only to dynamics which
evolve the 
random variable $Y_t$ i.e. the perturbed transition matrix is written
\begin{gather} 
\label{pertM}
M(\C,\C^\prime) = M_0(\C,\C^\prime) + e^\gamma M_1(\C,\C^\prime) +
e^{-\gamma} M_{-1}(\C,\C^\prime)\;,
\end{gather}
where $\gamma$, which measures the strength of the perturbation, is
small (and where $M(\C,\C)=0$). Then, to first order in $\gamma$, the
steady-state Master equation (\ref{p(C)}) requires that
\begin{gather} \label{1storder}
\sum_{\C^\prime} \left[ M(\C,\C^\prime) \frac{\partial
p(\C^\prime)}{\partial \gamma}
- M(\C^\prime,\C) \frac{\partial p(\C)}{\partial \gamma} \right] 
= 
-\sum_{\C^\prime} \left[ \frac{\partial M(\C,\C^\prime)}{\partial
\gamma} p(\C^\prime) -  
\frac{\partial M(\C^\prime,\C)}{\partial \gamma} p(\C) \right] 
\end{gather}
We can use (\ref{pertM}) in the rhs of this equation 
along with
the detailed balance conditions (\ref{DB2},\ref{DB3}) to write
\begin{gather} \label{pertp}
\sum_{\C^\prime} \left[ M(\C,\C^\prime) \frac{\partial
p(\C^\prime)}{\partial \gamma} 
- M(\C^\prime,\C) \frac{\partial p(\C)}{\partial \gamma} \right]
= -2\sum_{\C^{\prime}} \left[ M_1 (\C,\C^{\prime}) - M_{-1} 
(\C,\C^\prime) \right] p(\C^\prime)\;.
\end{gather}
By comparing equations (\ref{unpertr}) and (\ref{pertp}) we deduce
\begin{gather} \label{FRF}
2 r(\C) = \frac{\partial p(\C)}{\partial \gamma}\;.
\end{gather}
Therefore we can express the diffusion constant, given by (\ref{D}),
in terms of the response of the steady state to the
applied field:
\begin{gather} \label{D1}
\Delta = \sum_{\C,\C^\prime} \left[ M_1 (\C,\C^{\prime}) + M_{-1}
(\C,\C^\prime) \right] p(\C^\prime) + \sum_{\C,\C^\prime} \left[ M_1
(\C,\C^{\prime}) - M_{-1} (\C,\C^\prime) \right] \frac{\partial
p(\C^\prime)}{\partial \gamma} \;.
\end{gather} 
From (\ref{J})
the  expression of the response $\partial J/\partial \gamma$
of the current  to the perturbation
(\ref{pertM}), to first order in $\gamma$, is the same as the rhs
of (\ref{D1}). Thus,
\begin{gather} \label{ER}
\Delta = \frac{\partial J}{\partial \gamma}\;,
\end{gather}
which is the usual equilibrium Einstein relation.

Thus we have demonstrated an Einstein relation for all finite-state
stochastic models which obey detailed balance (in the absence of the
perturbing field).  Previously Einstein relations have been derived
for particular models in this class. For example, the symmetric
exclusion process \cite{ferrari85} and the zero-field repton model of
polymer dynamics\cite{vanLeeuwen91}.

%%%%%%%%%%%%%%%%%%%%%%%%%%%%%%%%%%%%%%%%%%%%%%%%%%%%%%%%%%%%%%%%%%%%%%%%%%%%%  
%%%% ******************************************************************* %%%%
%%%%%%%%%%%%%%%%%%%%%%%%%%%%%%%%%%%%%%%%%%%%%%%%%%%%%%%%%%%%%%%%%%%%%%%%%%%%%

\section{Einstein relation for nonequilibrium steady states} 
\label{sect:NSS}
In the absence of detailed balance, the structure of the steady state
is not known in general and it is not known how to formulate an
Einstein relation. 

For the case of detailed balance we showed in the previous section
that the Einstein relation resulted from the simple relationship given
in equation (\ref{FRF}), between $r(\C)$ and the response of $p(\C)$
to the applied field.  Inspired by this, in the following in cases where detailed
balance doesn't hold, our aim is to find a perturbation, parameterised
by a rate $\gamma$, such that
\begin{gather} \label{FRR}
2 r(\C) = \Omega \frac{\partial p(\C)}{\partial \gamma}\;,
\end{gather}
where $\Omega$ is a constant of proportionality. 
 
\subsection{Simple case of usual Einstein relation}

For a perturbation of the form
(\ref{pertM}),
it is clear by comparing (\ref{r(C)}) and (\ref{1storder})
that  (\ref{FRR}) holds 
for the  class of models  which
satisfies the condition
\begin{gather} \label{Jcond}
J = \sum_{\C^\prime} \left[ M_1 (\C^{\prime},\C) - M_{-1} (\C^\prime,\C)
\right]  \;,
\end{gather}
for all configurations $\C$. Using the condition (\ref{sumr(C)}), the
diffusion constant for this class of models can be written
\begin{gather} 
\Delta = \sum_{\C,\C^\prime} \left[ M_1 (\C,\C^{\prime}) + M_{-1}
(\C,\C^\prime) \right] p(\C^\prime) \;,
\end{gather}
which using the expression for $J$ (\ref{J})
yields the usual Einstein relation (\ref{ER}). 
Thus  (\ref{Jcond}) is a simple condition for 
the usual Einstein relation to hold for 
nonequilibrium systems.

We now show that the condition (\ref{Jcond}) holds for the boundary
driven symmetric exclusion process. This is a
stochastic model defined on a lattice of $N$ sites, where each site $i$
is either occupied by a particle (indicated by the variable 
$n_i$ taking value 1) or vacant (indicated by $n_i=0$). 
In one dimension, the dynamics in the bulk are
defined such that particles hop to 
the right (which we represent as the process $1 \, 0
\rightarrow 0 \, 1$) or the left (the process $0 \, 1 \rightarrow 1 \,
0$) with rate 1. At the
left-hand boundary site, particles are injected with rate $1-\rho_L$
and removed with rate $\rho_L$ and at the right-hand boundary site,
particles are removed with rate $\rho_R$ and injected with rate
$1-\rho_R$. Thus the system is in contact with boundary reservoirs
at densities $\rho_L$ at the left-hand boundary and $\rho_R$ at the
right-hand boundary. This forces a current through the system
(provided $\rho_L \neq \rho_R$) and in the steady state the density
profile is linear.

We define $Y_t$ such that it is increased (decreased) every time a particle is
added (removed) at the left-hand boundary, every time a particle hops
to the right (left), or every time a particle is removed (added) at
the right-hand boundary.
In the absence of the perturbation, the exact current is
\cite{sasamoto96} 
\begin{gather}
J = \rho_R - \rho_L \; .
\end{gather}
(Note that the current across a bond is $J/(N+1)$.)
To verify that (\ref{Jcond}) holds,
this is to be compared with
\begin{align}
\sum_{\C^\prime} \left[ M_1 (\C^{\prime},\C) - M_{-1} (\C^\prime,\C)
\right] =& (1-\rho_L) (1-n_1) -\rho_L n_1
+ \rho_R n_N - (1-\rho_R) (1-n_N)
\nonumber \\
&+ \sum_{i=0}^{N-1} \left[
n_i(1-n_{i+1}) - (1-n_i)n_{i+1} \right] \; .
\end{align} 
For any configuration $\C=\{ n_i \}$, the terms in the sum over $i$
cancel except near the boundaries, and one obtains
\begin{gather}
\sum_{\C^\prime} \left[ M_1 (\C^{\prime},\C) - M_{-1} (\C^\prime,\C)
\right] = \rho_R - \rho_L \;,
\end{gather}
for every configuration $\C$. Therefore the condition (\ref{Jcond})
holds and the Einstein relation (\ref{ER}) follows.

The class of models for which (\ref{Jcond})
holds
includes the boundary driven symmetric zero-range process
and other models where detailed balance is broken by the boundary
dynamics i.e.  models which would satisfy detailed balance 
if periodic boundary conditions were imposed.

%%%%%%%%%%%%%%%%%%%%%%%%%%%%%%%%%%%%%%%%%%%%%%%%%%%%%%%%%%%%%%%%%%%%%%%%%%%%%  
%%%% ******************************************************************* %%%%
%%%%%%%%%%%%%%%%%%%%%%%%%%%%%%%%%%%%%%%%%%%%%%%%%%%%%%%%%%%%%%%%%%%%%%%%%%%%%

\subsection{Einstein relation for the ASEP}
\label{subsect:ASEP}
We now consider the asymmetric simple exclusion process (ASEP).  In
one dimension the model, with periodic boundary conditions, is defined
such that particles hop only to the right with rate 1 (i.e. the
process $1 \, 0 \rightarrow 0 \, 1$).  In the steady state, all
configurations occur with equal likelihood, so $p(\C)$ is given by the
binomial coefficient,
\begin{gather} \label{ss}
p(\C) =  \left[ \binom{N}{M} \right]^{-1} \equiv Z_{N,M}^{-1} \;,
\end{gather}
for a ring of $N$ sites containing $M$ particles.  For the ASEP
neither detailed balance nor the condition (\ref{Jcond}) are
satisfied. However, by considering a perturbation which creates and
annihilates particles, an Einstein relation can be derived in the
following way.

We define $Y_t$ such that it is incremented every time  any particle
performs a hop, so the current (\ref{J}) is given by
\begin{gather}
J=NZ_{N,M}^{-1} Z_{N-2,M-1}\;,
\end{gather}
and using equation (\ref{r(C)}), $r(\C)$ must satisfy
\begin{gather} \label{unpert}
\sum_{\C^\prime} \left[ M(\C,\C^\prime) r(\C^\prime) - 
M(\C^\prime,\C) r(\C) \right] = (NZ_{N,M}^{-1} Z_{N-2,M-1} - l_{10})\,p(\C)\;,
\end{gather}
where $l_{10}=\sum_{i=1}^N n_i (1-n_{i{+}1})$ is the number of $1 \, 0$ 
configurations across all bonds in the system configuration $\C$. 

Now consider the perturbed transition matrix for the ASEP,
\begin{gather}
M(\C,\C^\prime) = M_1(\C,\C^\prime) + M_\gamma(\C,\C^\prime)\;,
\end{gather} 
where the elements of $M_1(\C,\C^\prime)$ describe the unperturbed ASEP
and the elements of $M_\gamma(\C,\C^\prime)$ describe
the following processes that occur at all pairs of nearest neighbour sites
\begin{gather}
1 \, 1 \; \stackrel{\gamma\epsilon_1}{\longrightarrow}  \; 1 \, 0  \;,\qquad \qquad 
1 \, 1 \; \stackrel{\gamma\epsilon_3}{\longrightarrow}  \; 0 \, 1  \;,\nonumber \\ \label{pertG}
0 \, 0 \; \stackrel{\gamma\epsilon_2}{\longrightarrow}  \; 1 \, 0  \;,\qquad \qquad 
0 \, 0 \; \stackrel{\gamma\epsilon_4}{\longrightarrow}  \; 0 \, 1  \;,
\end{gather}
where $\gamma$ is a rate and $\{\epsilon_i\}$ measure the relative
strengths of the four processes.  Since this perturbation creates and
annihilates particles, the different particle sectors
(i.e. configurations with the same number of particles) are now
connected by the dynamics. This modifies the steady state: if we
define $P(M)$ to be the (normalised) probability that the system is in
the sector containing $M$ particles, then, for $\gamma \to 0$, the
probability that the system is in the configuration $\C$ in the
$M$-particle sector, $p_M(\C)$, can be written as
\begin{gather}
p_M(\C) = P(M) p(\C) \; 
\end{gather}
where $p(\C)$ is the steady state probability in the absence of the
perturbation. 
$P(M)$ is determined by the following balance
condition 
with respect to transitions between particle sectors in the limit
$\gamma \to 0$: 
\begin{gather} \label{DBP(M)}
(\epsilon_1 + \epsilon_3) \, \langle 1\,1 \rangle_{M+1} \, P(M+1) =
(\epsilon_2+\epsilon_4) \, \langle 0\,0 \, \rangle_M P(M) \;,
\end{gather}
where $\langle \cdot \rangle_M$ represents a correlation function
calculated with the respect to the steady state (\ref{ss}) in the
$M$-particle sector. Since in the limit $\gamma \to 0$ all
configurations within each sector are equally likely, these
correlation functions are easily found to be $\langle 1\,1
\rangle_{M+1} = Z_{N,M+1}^{-1} Z_{N-2,M-1}$ and $\langle 0\,0 \,
\rangle_M = Z_{N,M}^{-1} Z_{N-2,M}$. Consequently the solution of
(\ref{DBP(M)}) is readily obtained as
\begin{gather}
P(M) = \Lambda_N^{-1}
\left(\frac{\epsilon_2+\epsilon_4}{\epsilon_1+\epsilon_3}\right)^M Z_{N,M} \,
Z_{N-2,M-1} \;,
\end{gather}
leading to the modified steady
state $p_M(\C)$ given by
\begin{gather} \label{p_M(C)}
p_M(\C) = \left(\frac{\epsilon_2+\epsilon_4}{\epsilon_1+\epsilon_3}\right)^M
\Lambda_N^{-1} Z_{N-2,M-1} \; ,
\end{gather}
where $\Lambda_N$ is a  normalisation, fixed by the requirement
that $\sum_M P(M) = 1$.

For the ASEP, (\ref{FRR}) is shown to hold as follows.
By (\ref{1storder}),
\begin{align} \label{pert}
\sum_{\C^\prime} \left[ M(\C,\C^\prime) \frac{\partial p_M(\C^\prime)}{\partial \gamma}
- M(\C^\prime,\C) \frac{\partial p_M(\C)}{\partial \gamma} \right] =&
(\epsilon_1+\epsilon_3) l_{11} p_M(\C) + (\epsilon_2+\epsilon_4) l_{00}
p_M(\C) 
\nonumber \\ 
& \quad - \epsilon_1 l_{10} p_{M+1}(\C) - \epsilon_3 l_{01} p_{M+1}(\C)
\nonumber \\  
& \quad - \epsilon_2 l_{10} p_{M-1}(\C) - \epsilon_4 l_{01} p_{M-1}(\C)
\; .
\end{align}
The rhs is simplified by noting that $l_{10} = l_{01}$ and
$l_{11} + l_{10} = M$ and $l_{00}+l_{10}=N-M$, using (\ref{p_M(C)}),
and  exploiting the identity $Z_{N,M} = Z_{N-2,M-2} + 2Z_{N-2,M-1} +
Z_{N-2,M}$. Thus, it is straightforward to show that the rhs of
(\ref{pert}) is
proportional to the rhs of equation (\ref{unpert}), therefore $r(\C)$ is
proportional to $\partial p_M(\C) / \partial \gamma$ and the constant
of proportionality $\Omega$ is
\begin{gather} \label{Omega}
\Omega =  \left( \frac{\epsilon_2+\epsilon_4}{\epsilon_1+\epsilon_3}
\right)^{-M} \Lambda_N Z_{N,M}^{-2} \;.
\end{gather}
Hence the diffusion constant (\ref{D}) can be expressed in terms of
the response of the steady state to the perturbation, leading to an
Einstein relation of the form
\begin{gather} \label{NER}
\Delta = J+\Omega\frac{\partial J}{\partial \gamma} \;.
\end{gather}

Thus for the ASEP we have shown that an Einstein relation
(\ref{NER}) holds which expresses the diffusion constant
as the current $J$ plus the response of the current to  perturbations
(\ref{pertG}) that create and annihilate particles.
However it should be noted that not all perturbations
that create and annihilate particles lead to 
(\ref{FRR}). For example creating and annihilating particles
at a site regardless of the occupation of
neighbouring  sites does not satisfy (\ref{FRR}).

It can be shown that
a relation of the form (\ref{FRR})
holds for several variations of the ASEP under suitable perturbations
which we now describe.
\begin{description}
\item[a) Partially asymmetric exclusion:]
in the partially asymmetric exclusion process
(PASEP) particles
hop to the left with rate $q$ and to the right with rate $p$
(if the target sites are empty).
This model can be treated, as above,  by introducing the perturbation
(\ref{pertG}), leading to an Einstein relation of the form
\begin{gather} \label{partialNER} 
\Delta = \frac{p+q}{p-q}J+(p-q)\Omega\frac{\partial J}{\partial \gamma} \;,
\end{gather}
where the current $J=(p-q)Z_{N,M}^{-1}Z_{N-2,M-1}$ and $\Omega$
is given by (\ref{Omega}).

\item[b) PASEP in higher dimension:]
an Einstein relation holds 
for the PASEP
in any dimension,
on a hypercubic lattice
with fully periodic boundary conditions. 
To illustrate this, consider a two dimensional
square lattice on which particles hop up and to the right with rate
$p$, and down and to the left with rate $q$. 
The random variable $Y_t$ is defined such that $Y_t$
increases whenever a particle hops up or to the right and decreases
whenever a particle hops down or to the left. By considering a
perturbation
\begin{gather}
1 \, 1 \; \stackrel{\gamma}{\longrightarrow}  \; 1 \, 0  \;,\qquad \qquad 
1 \, 1 \; \stackrel{\gamma}{\longrightarrow}  \; 0 \, 1  \;,\qquad \qquad 
0 \, 0 \; \stackrel{\gamma}{\longrightarrow}  \; 1 \, 0  \;,\qquad \qquad 
0 \, 0 \; \stackrel{\gamma}{\longrightarrow}  \; 0 \, 1  \;,\nonumber
\end{gather}
applied to horizontal pairs of nearest neighbour sites, and
\begin{gather}
\shortstack{ 1 \\ 1 } \; {\stackrel{\gamma}{\longrightarrow}\atop} \;
\shortstack{ 0 \\ 1 } \;{,\atop}\qquad \qquad 
\shortstack{ 1 \\ 1 } \; {\stackrel{\gamma}{\longrightarrow}\atop}  \; 
\shortstack{ 1 \\ 0 } \;{,\atop}\qquad \qquad 
\shortstack{ 0 \\ 0 } \; {\stackrel{\gamma}{\longrightarrow}\atop}  \; 
\shortstack{ 0 \\ 1 } \;{,\atop}\qquad \qquad 
\shortstack{ 0 \\ 0 } \; {\stackrel{\gamma}{\longrightarrow}\atop}  \; 
\shortstack{ 1 \\ 0 } \;{,\atop} \nonumber 
\end{gather}
applied to vertical pairs of nearest neighbour sites,
an Einstein relation of the form (\ref{partialNER}) can be derived,
where $\Omega$ is given by $\Omega = \Lambda_N Z_{N,M}^{-2}$. 

Further, if the dynamics are such that particles hop up and
down with the same rate 1, and to the right with rate $p$ and to the
left with rate $q$, then, again, the Einstein relation
(\ref{partialNER}) can be derived. This is
achieved by defining $Y_t$ such that $Y_t$ is 
increased when a particle hops to the right and decreased when a
particle hops to the left (i.e. $Y_t$ is unaltered by hops up or
down). The perturbation (\ref{pertG}) is then applied to pairs of
nearest neighbour sites
only along the axis of asymmetric hopping.

\item[c) A marked bond:]
if $Y_t$ is incremented only whenever a particle hops across a single,
specified bond (the `marked' bond), then a perturbation applied across this
bond only and defined by the processes
\begin{gather}
1 \, 1 \; \stackrel{\gamma}{\longrightarrow}  \; 1 \, 0 \;,
\qquad \qquad  
0 \, 0 \; \stackrel{\gamma}{\longrightarrow}  \; 1 \, 0 \;,
\qquad \qquad 
0 \, 1 \; \stackrel{\gamma}{\longrightarrow}  \; 1 \, 0 \;,
\end{gather} 
yields  a relation of the form (\ref{NER}) with $\Omega = \Lambda_N
Z_{N,M}^{-2}$.

\end{description}

Finally we note that for symmetric hopping, which satisfies detailed
balance, the kinds of
perturbations given in this section (i.e. those linking particle
sectors without coupling to the original dynamics) do not satisfy the
condition (\ref{FRR}) -- rather, the appropriate perturbation couples
only to the hopping dynamics, as discussed in section \ref{sect:DB}.

%%%%%%%%%%%%%%%%%%%%%%%%%%%%%%%%%%%%%%%%%%%%%%%%%%%%%%%%%%%%%%%%%%%%%%%%%%%%%  
%%%% ****************************************************************** %%%%%
%%%%%%%%%%%%%%%%%%%%%%%%%%%%%%%%%%%%%%%%%%%%%%%%%%%%%%%%%%%%%%%%%%%%%%%%%%%%%

\subsection{Einstein relation for dimer diffusion}
In a similar way to the ASEP discussed in the previous
subsection,
we are also able to obtain an Einstein relation for (reconstituting)
dimer diffusion: the process
\begin{gather} \label{dimerD}
1 \, 1 \, 0 \; {\longrightarrow}  \; 0 \, 1 \, 1 \;,
\end{gather}
which occurs at all nearest neighbour triples of sites with rate $1$.
We assume the density is greater than one half so that
there are no configurations where all particles are isolated.
Again, in the steady state of this model, all configurations are
equally likely; we define $Y_t$ such that it increases by one every time
the process (\ref{dimerD}) takes place (anywhere on the lattice), 
hence the current is
given by $J = N Z_{N,M}^{-1} Z_{N-3,M-2}$. 
The Einstein relation ({\ref{NER}) 
is derived for a perturbation defined by the processes
\begin{gather}
1 \, 1 \, 1 \; \stackrel{\gamma}{\longrightarrow}  \; 1 \, 1 \, 0 \;,
\qquad \qquad  
1 \, 0 \, 0 \; \stackrel{\gamma}{\longrightarrow}  \; 1 \, 1 \, 0 \;,
\qquad \qquad 
1 \, 0 \, 1 \; \stackrel{\gamma}{\longrightarrow}  \; 1 \, 1 \, 0 \;.
\end{gather} 
As before, the steady state distribution is modified to accommodate a
balance condition with respect to transitions between particle sectors
c.f.  (\ref{DBP(M)}). Using the same reasoning as that applied to the
ASEP, $r(\C)$ is shown to be proportional the the response of the
steady state to the perturbation, where the constant of
proportionality $\Omega$ is
\begin{gather}
\Omega = \Lambda_N Z_{N{-}1,M{-}1}^{-1} Z_{N,M}^{-1} \;,
\end{gather}
and the diffusion constant and the response of the current to the
perturbation are related by the Einstein relation given in equation
(\ref{NER}).  

\subsection{Einstein relation for the boundary driven ASEP}
The boundary driven asymmetric simple exclusion process 
also satisfies an Einstein relation. In this model, 
particles can hop only to the right, and the multiple occupancy of any
site is still forbidden. At the left-hand boundary site particles 
are injected with rate $\alpha$ and 
at the right-hand boundary site they are removed with rate $\beta$.
In this case, $Y_t$ is incremented every time a particle
is added at the left-hand boundary site.
Along the special line $\alpha+\beta=1$, all
configurations in the same particle sector are equally likely in the steady
state, and it is known that \cite{derrida95}
\begin{gather}
r(\C) = \alpha(1-\alpha)\frac{dp(\C)}{d\alpha} \; ,
\end{gather}
is a solution of (\ref{r(C)}). This perturbation corresponds to a
small increase in $\alpha$ away from the line $\alpha+\beta=1$, thereby changing slightly the density of
the left-hand boundary reservoir.
Hence,  one finds an Einstein relation of the form
\begin{gather}
\Delta = (2\alpha -1)J + 2 \alpha (1-\alpha) \frac{\partial
J}{\partial \alpha}\; . 
\end{gather}

%%%%%%%%%%%%%%%%%%%%%%%%%%%%%%%%%%%%%%%%%%%%%%%%%%%%%%%%%%%%%%%%%%%%%%%%%%%%%  
%%%% ******************************************************************* %%%%
%%%%%%%%%%%%%%%%%%%%%%%%%%%%%%%%%%%%%%%%%%%%%%%%%%%%%%%%%%%%%%%%%%%%%%%%%%%%%

\section{Conclusion}
\label{sect:conc}
In this work we have investigated the existence of Einstein relations
for stochastic, lattice models.  Within the framework reviewed in
section~\ref{sect:DA} we have shown that an Einstein relation will
result if one  can express the quantities $r(\C)$ as the response of
the steady state probabilities $p(\C)$ to a perturbation.  For the case of detailed balance
we showed that this can always be done through the perturbation
(\ref{pertM}) which yields the usual Einstein relation (\ref{ER}).
Further for a class of nonequilibrium steady states where
(\ref{Jcond}) holds one again has the usual Einstein relation.  This
class includes models where detailed balance is broken only by the
boundary dynamics such as boundary-driven symmetric exclusion
\cite{bertini01,bertini02}.

Turning to systems where detailed balance is lacking even under
periodic boundary conditions we have found nonequilibrium
generalisations of the Einstein relation for some specific models. In
these cases the perturbation creates and annihilates particles,
breaking the conservation law of the unperturbed dynamics.

The models for which we succeeded in finding an Einstein relation had
the simplifying property that in each particle sector all
configurations are equally likely.  It remains a challenge to
establish whether Einstein relations hold for more general
nonequilibrium steady states. Natural starting points would be steady
states that have a simple structure such as the one dimensional KLS
model \cite{katz84}.

\noindent {\bf Acknowledgements}
We thank  D. Mukamel for helpful suggestions.
MRE also  thanks B. Derrida and G. Sch\"utz for
many discussions on this topic.

\null

%\bibliographystyle{/Home/thanney/myreps/Bib_styles/elsart-num}
%\bibliography{/Home/thanney/myreps/References/refs}

\end{document}